\documentclass[iop,apjl,tighten]{emulateapj}
\usepackage{graphicx,enumitem}

\usepackage{color}

\newcommand{\Msun}{{M_\odot}}
\usepackage{soul}   

\shorttitle{The Milky Way Halo in Action Space}
\shortauthors{Myeong et al.}

\begin{document}

\title{The Milky Way Halo in Action Space}

\author{G.~C. Myeong$^{1}$, N.~W. Evans$^1$, V. Belokurov$^1$,
  J.~L. Sanders$^1$ and S.~E. Koposov$^{1,2}$}

\affiliation{$^1$Institute of Astronomy, University of Cambridge,
  Madingley Road, Cambridge CB3~0HA, United Kingdom\\ $^2$McWilliams
  Center for Cosmology, Department of Physics, Carnegie Mellon
  University, 5000 Forbes Avenue, Pittsburgh PA 15213, USA}

\begin{abstract}
  We analyse the structure of the local stellar halo of the Milky Way
  using $\sim$ 60\,000 stars with full phase space coordinates
    extracted from the SDSS--{\it Gaia} catalogue. We display stars in
    action space as a function of metallicity in a realistic
    axisymmetric potential for the Milky Way Galaxy.  The metal-rich
    population is more distended towards high radial action $J_R$ as
    compared to azimuthal or vertical action, $J_\phi$ or $J_z$.  It
    has a mild prograde rotation $(\langle v_\phi \rangle \approx 25$
    km\,s$^{-1}$), is radially anisotropic and highly flattened with
    axis ratio $q \approx 0.6 - 0.7$. The metal-poor population is
    more evenly distributed in all three actions. It has larger
    prograde rotation $(\langle v_\phi \rangle \approx 50$
    km\,s$^{-1}$), a mild radial anisotropy and a roundish morphology
    ($q\approx 0.9$). We identify two further components of the halo
    in action space. There is a high energy, retrograde component that
    is only present in the metal-rich stars. This is suggestive of an
    origin in a retrograde encounter, possibly the one that created
    the stripped dwarf galaxy nucleus, $\omega$Centauri.  Also visible
    as a distinct entity in action space is a resonant component,
    which is flattened and prograde. It extends over a range of
    metallicities down to [Fe/H] $\approx -3$. It has a net outward
    radial velocity $\langle v_R \rangle \approx 12$ km\,s$^{-1}$
    within the Solar circle at $|z| <3.5$ kpc. The existence of
    resonant stars at such extremely low metallicities has not been
    seen before.
\end{abstract}


\section{Introduction}

Samples of halo stars near the Sun provide us with accessible
documentation on the early history of the Galaxy.  But, like medieval
palimpsests, the manuscript pages have been overwritten. Dissipative
events, such as the assembly of the Galactic disk, and dissipationless
processes, such as phase-mixing in the time-evolving Galactic
potential, make the text difficult to read.  Nonetheless, the spatial
distribution, kinematics and chemistry of local halo stars provides us
with important evidence on the nature and timescale of events in the
early history of the Galaxy, if we could but decode and interpret it.

Actions are invariant under slow changes~\citep[e.g.,][]{Go80}. They
have often been suggested as the natural coordinates for galactic
dynamics~\citep{Bi82,Bi87}, in which of course the potential is
evolving in time (though possibly not slowly).  One of the advances
over the last few years has been the development of fast numerical
methods to compute actions in general axisymmetric
potentials~\citep{Bi12,Bo15,Sa16}. For the first time, this allows
  the study of the local halo in action space using realistic Galactic
  potentials comprising disks (both stellar and gas), halo and
  bulge~\citep{Mc17}. The only other work known to us that displays
  the local halo in action space is the pioneering paper of
  \citet{Ch00}. These authors worked with a much smaller sample of
  stars ($\sim 1000$) and out of necessity used a global St{\"a}ckel
  potential as a model of the Galaxy, as their work predated fast
  numerical action evaluators.

In this {\it Letter}, we map the halo stars in action space using
  a new dataset, the SDSS--{\it Gaia} catalogue. This has six-dimensional 
  phase space information for 62\,133 halo stars, an order of magnitude
  larger than previous studies of the local halo. We use this new
  dataset, coupled with the recent advances in action evaluation, to
  provide new maps of the local halo in action space, which
  graphically illustrate the dichotomy between the metal-rich and
  metal-poor stars.  We identify two new components of the
  halo. First, there is a high energy retrograde component that is
  limited to the metal-rich stars. Although retrograde stars have been
  identified before~\citep{Ma12,He17}, we show here that they are
  largely restricted to metallicities in the range $-1.9 < $ [Fe/H] $<
  -1.3$. Second, we provide strong evidence for the existence of a
  resonant component. It is present across all metallicities, and it
  has a strong spatial dependence, which satisfies the characteristics
  of a dynamically induced resonance. This resonance may be linked to
  a well-known resonance in disk stars, which causes the Hercules
  Stream. The presence of resonant stars at such low metallicities has
  not been seen before. We conclude with a discussion of how these
findings are related to fundamental events in the Galaxy's early
history.

\begin{figure*}
    \includegraphics[width=180mm]{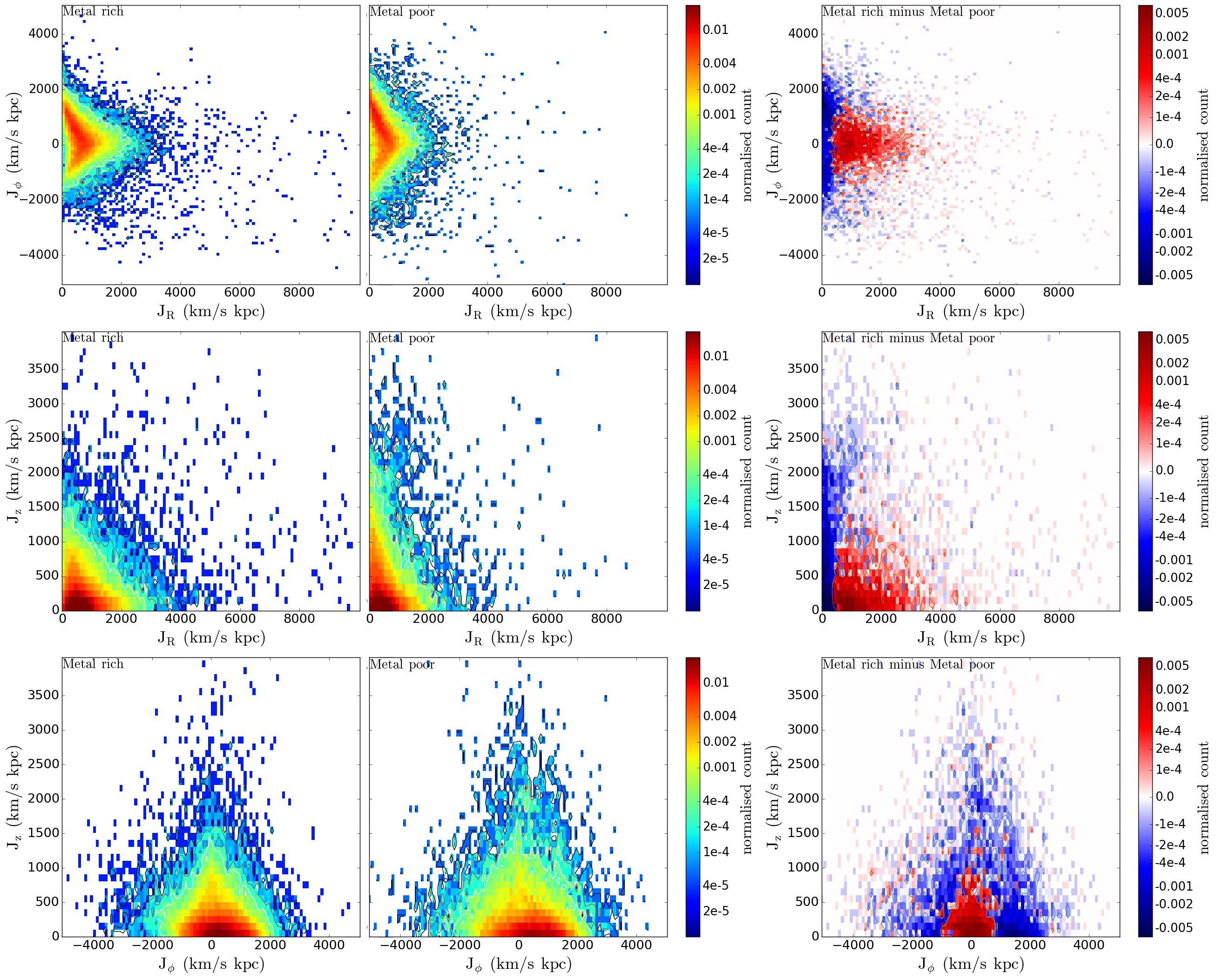}
  \caption{Histograms of the stellar halo in action space
    ($J_R,J_\phi$), ($J_R,J_z$) and ($J_\phi,J_z$) split into
    metal-rich (left column, $-1.6<$ [Fe/H] $< -1.1$) and metal-poor
    (middle column, $-2.9 <$ [Fe/H] $< -1.8$). The right column displays
     the difference, with red showing an excess of metal-rich, blue an
    excess of metal-poor stars. Notice (i) the metal-rich stars are
    tightly clustered around $J_\phi \approx 0$ and are much more
    extended in $J_R$, and (ii) the metal-poor stars have prograde
    rotation ($\langle J_\phi \rangle >0$) and a more isotropic
    distribution in action space.
\label{fig:actionspace}}
\end{figure*}

\section{The SDSS--{\it Gaia} Catalogue}

The SDSS--{\it Gaia} catalogue was constructed by Sergey Koposov and
contains all SDSS stars down to $r=20.5$.  The catalogue was made by
recalibrating the Sloan Digital Sky Survey (SDSS) astrometric
solution, and then obtaining proper motions from positions in the {\it
  Gaia} Data Release 1 (DR1) Source catalogue~\citep{Gaia1,Gaia2} and
their recalibrated positions in SDSS~\citep[see e.g.,][for more
  details] {De17,DeBoer18}. The individual SDSS--{\it Gaia} proper
motions have statistical errors typically $\sim 2$ mas yr$^{-1}$, or
$\sim 9.48 D$ km s$^{-1}$ for a star with heliocentric distance $D$
kpc.  We work here mainly with the subsample of main sequence stars
and blue horizontal branch stars with available spectroscopic
measurements, in heliocentric distances of $\lesssim 10$ kpc.  The
spectroscopic measurements such as radial velocities and metallicities
are obtained from SDSS DR9 spectroscopy~\citep{Ahn12} or the
crossmatch with LAMOST DR3~\citep{Lu15}.  Finally, photometric
parallaxes for stars such as main-sequence turn-offs (MSTOs) or blue
horizontal branch stars (BHBs) can be added using the formulae in
\citet{Iv08} and \citet{De11b} to give samples with the full
six-dimensional phase space coordinates.

The SDSS--{\it Gaia} catalogue currently provides the most extensive
catalogue of halo stars with positions and kinematics, albeit without
an easily calculable selection function. This means that some stellar
orbits are not properly represented (such as stars with low-$J_R$ and
low-$J_z$ orbits in small Galactocentric radius), but this will not
change the gross morphological features we describe. The actions and
energy of each star are estimated using the numerical method of
\citet{Bi12} and \citet{Sa16} together with the potential of
\citet{Mc17}. The latter also provides the circular speed at the
  Sun as 232.8 kms$^{-1}$, whilst for the Solar peculiar motion we use
  the most recent value from \citet{Sc10}, namely $(U,V,W) = (11.1,
  12.24, 7.25)$ kms$^{-1}$. Quality cuts and the disk-halo
separation employed by \citet{My18}, together with an additional cut
on distance error $< 2.5$ kpc, are used to clean the sample. The mode
of the distance error is $\sim 0.32$ kpc.  As a result, we obtain a
sample of 62\,133 halo stars (61\,911 MSTO stars and 222 BHB stars;
59\,811 stars with SDSS DR9 and 2\,322 stars with LAMOST DR3
spectroscopy) with full six-dimensional phase space information,
actions and energy~\citep[see][for more details on the cuts and the
  sample]{Wi17,My18}.  The metallicity distribution of the halo stars
shows evidence of at least two subpopulations~\citep[see e.g., Figure
  1 of][]{My18}. A Gaussian Mixture model (GMM) from the {\tt
  Scikit-learn} \citep{Pe11}\footnote{http://scikit-learn.org} based
on metallicity suggests the subdivision of the halo sample at [Fe/H]
$\approx -1.67$, with 37\,670 metal-rich and 24\,463 metal-poor halo
stars.

\begin{figure}
    \includegraphics[width=90mm]{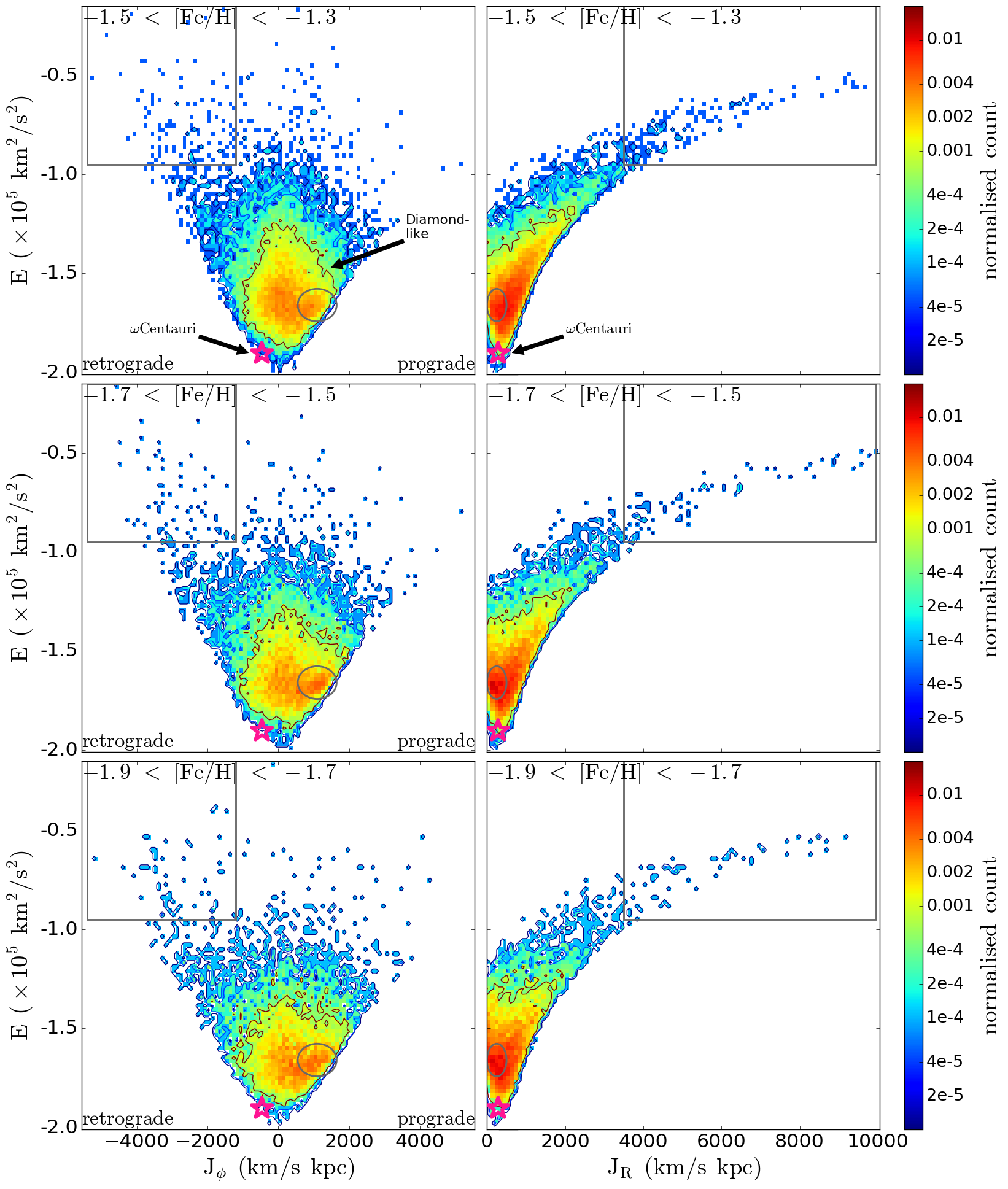}
    \includegraphics[width=90mm]{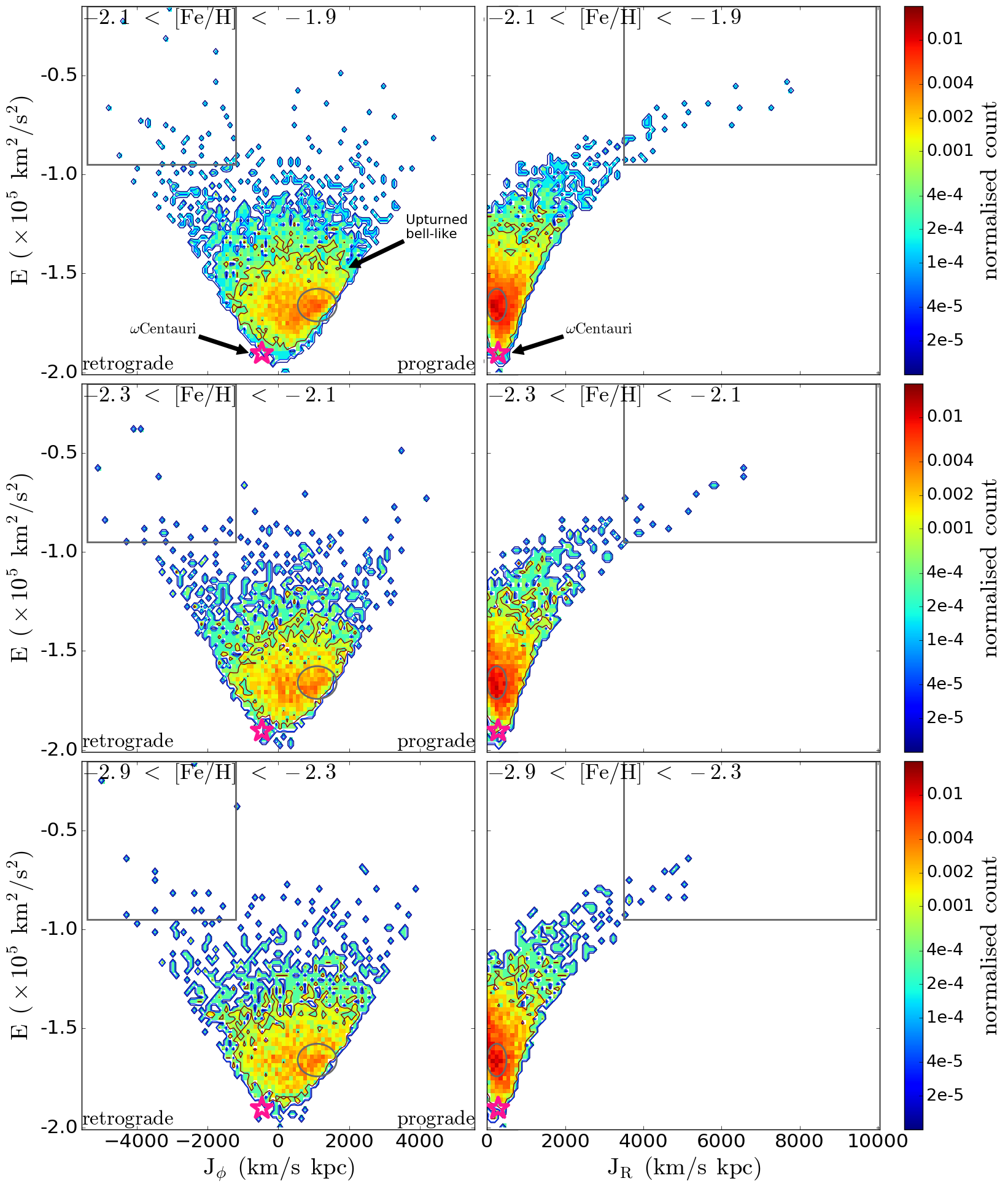}
    \caption{The distribution of halo stars in energy-action space or
      ($J_\phi,E$) and ($J_R, E$) space, split according to six
      metallicity bins from $-2.9 <$ [Fe/H] $< -1.3$. As we move from
      the most metal-rich to the most metal-poor, notice that (i) the
      distribution in radial action becomes more compact as the tail
      melts away by [Fe/H] $\approx -2.0$, (ii) the diamond-like shape
      of the contours in ($J_\phi,E$) changes gradually into an
      upturned bell-like shape, and (iii) the high energy, retrograde
      stars (marked by rectangular boxes in the left panels) and
      the high energy, eccentric stars (boxes in the right panels)
    gradually disappear by [Fe/H] $\approx -1.9$, (iv) there is a
    distinct prograde component at $J_\phi\approx 1100$, $J_R \approx$
    150 km\,s$^{-1}$ kpc, $E \approx -1.6$ km$^2$s$^{-2}$ (marked by
    ellipses), which is present at all metallicities. The location of
    $\omega$Centauri is shown by a pink star.
\label{fig:actspace}}
\end{figure}

\section{Characteristics of the Halo in Action Space}

\subsection{The Rich and the Poor}

We begin by showing the halo stars in action space ($J_R,J_\phi,
J_z$). For illustration, we show a metal-rich ($-1.6<$ [Fe/H] $<
-1.1$) and metal-poor ($-2.9 <$ [Fe/H] $< -1.8$) sample, together with
the difference between them in Fig.~\ref{fig:actionspace}. The
stratification on linear combinations of the actions suggested by
\citet{Wi15} and \citet{Po15} is discernible in the triangular shapes
of the contours in the left and centre panels. However, the metal-rich
sample is clearly much more distended toward high $J_R$ as compared
to $J_\phi$ or $J_z$. This is most evident in the difference plot in
the rightmost plot of Fig.~\ref{fig:actionspace}, in which the red is
preponderant at large values of $J_R$, and at low values of $J_\phi$
and $J_z$. The presence of abundant high eccentricity stars indicates
that the population is radially anisotropic, whilst the narrow spread
in $J_z$ suggests the population is also flattened. In contrast, the
metal-poor sample is distributed more equally in all three
actions. The blue is preponderant at high $J_z$ and reaches out to
larger values of $J_\phi$ in the rightmost column of
Fig.~\ref{fig:actionspace}. This suggests it is rounder and has a mild
net prograde rotation. Although correlations between kinematics and
metallicities of halo stars have been reported
before~\citep[e.g.,][]{Ch00,De11a,Be12,Ha13}, our pictures of the
  halo in action space provide a dramatic illustration of the
dichotomy between the metal-rich and metal-poor stars.

\begin{figure*}
\begin{center}
    \includegraphics[width=180mm]{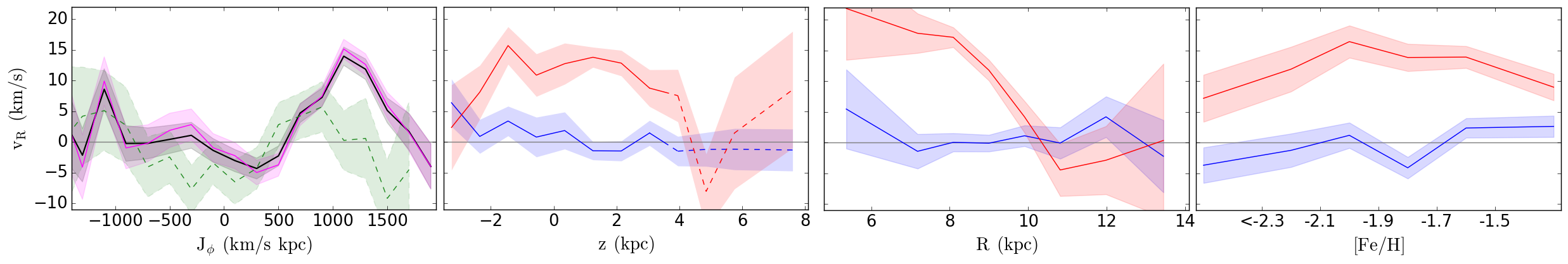}
\end{center}
  \caption{The Outward Moving Stars. Leftmost: The behaviour of the
    mean radial velocity $\langle v_R \rangle$ as a function of
    $J_\phi$ for stars with $|z| <3.5$ kpc (magenta), as compared to
    the whole sample (black). All show a signal of positive $\langle
    v_R\rangle$ at $J_\phi\approx 1100$ km\,s$^{-1}$ kpc. However, the
    high latitude stars ($|z| > 3.5$ kpc) shown in green do not
    contribute to the signal. Center and Rightmost: The panels show
    the trends in $\langle v_R \rangle$ of the group selected by
    Gaussian fitting in Sec.~\ref{sec:td} (red) and the rest (blue)
    against the Galactic position and metallicity. The selected stars
    with $|z| < 3.5$ kpc (full lines) are the main contributors to the
    outward moving population. There is no clear correlation with
    metallicity [Fe/H], but the stars are mainly found interior to the
    Solar circle.  For all panels, the shaded region around each line
    indicates the corresponding standard error.
    \label{fig:thirdcomp}}
\end{figure*}

Clearer details can be uncovered by slicing the halo into a sequence
of smaller metallicity bins, as in Fig.~\ref{fig:actspace}.  For the
most metal-rich stars, the contours in $(J_\phi,E)$ space at high
energy are noticeably ``pointy''. There is a tight distribution around
$J_\phi \approx 0$ km\,s$^{-1}$ kpc, again indicating the presence of
many stars moving on nearly radial orbits. In general, the contours in
the metal-rich panel ([Fe/H] $>-1.5$) are diamond-like in $(J_\phi,E)$
space, while they resemble an up-turned bell for the metal-poor([Fe/H]
$<-2.1$). Related to this, the metal-rich sample in $(J_R,E)$ space is
skewed strongly toward high $J_R$, while in contrast, the metal-poor
sample shows much less spread toward high $J_R$. In fact, in the more
metal-poor panels ($-2.9 <$ [Fe/H] $<-1.9$), the bulk of the
distribution (coloured red) shows the reverse trend of decreasing
$J_R$ with increasing energy $E$.

The metal-rich stars comprise a radially anisotropic and flattened
population. The highest energy and most metal-rich stars are strongly
retrograde, but the bulk of the population is at lower energies and
shows mild prograde rotation.  The metal-poor stars form a rounder
population. This is also suggested by the broader distribution in $J_z$ in
Fig.~\ref{fig:actionspace}. The kinematics are more isotropic, and there
is significant prograde rotation.  To obtain an idea of the
three-dimensional shape, we study their kinematics.  Properly
speaking, we should decompose the distribution into components in
action space, much as \citet{Be18} do in velocity space.  Here, we
will simply assume that contamination is mild in the extremal
metallicity bins.  In the most metal-rich bin, the rotational velocity
$\langle v_\phi \rangle = 25$ km\,s$^{-1}$, whilst the velocity
dispersion tensor is radially anisotropic with
$(\sigma_R,\sigma_\phi,\sigma_z) = (155,77,88)$ km\,s$^{-1}$. So, the
ratio of the horizontal to the vertical components of the kinetic
energy tensor is $\approx 4.0$. This is equal to the ratio of
horizontal to vertical components of the potential energy tensor via
the virial theorem. It can be used to calculate the intrinsic shape,
as advocated in \citet{Ag12}. Using their Figure 1, we see that the
axis ratio of the density contours of the metal-rich population is $q
\approx 0.6$ - $0.7$, depending only modestly on the radial density
profile. By contrast, in the most metal-poor bin, the rotational
velocity is $\langle v_\phi \rangle = 49$ km\,s$^{-1}$, whilst the
velocity dispersion tensor is close to isotropic with
$(\sigma_R,\sigma_\phi,\sigma_z) = (125,114,110)$ km\,s$^{-1}$. This
gives an axis ratio for the population of $q \approx 0.9$, so that the
density contours are very round.  These calculations assume that the
total Galactic potential (stars plus dark matter) is spherical. Any
flattening in the total potential will result in the computed axis
ratios becoming flatter.

\subsection{The Retrograde Stars}

Marked by rectangular boxes in the panels of Fig.~\ref{fig:actspace}
are the general location of the high energy (e.g., $E>-1.1\times10^5$
km$^2$s$^{-2}$), retrograde ($J_\phi < 0$) stars. The box is
well-populated in the metal-rich panels (e.g., [Fe/H] $>-1.9$), but
sparsely populated in the metal-poor.  These metal-rich, high energy
stars have large radial action $J_R$ as well, indicating a highly
eccentric, retrograde population. This trend diminishes with
decreasing metallicity and the metal-poorer panels (e.g., [Fe/H]
$<-1.9$) show a more evenly balanced $J_\phi$ distribution of high
energy stars.  In addition, irrespective of the sign of $J_\phi$,
there are noticeably more stars with very high energy (e.g.,
$E>-0.75\times10^5$ km$^2$ s$^{-2}$) in the metal-rich panels of
Fig.~\ref{fig:actspace} than the metal-poor. The overdensity of
retrograde high energy stars is evidence of a considerable
(retrograde) merger or accretion event in the
past~\citep[e.g.,][]{Qu86,No89}.

\subsection{The Resonant Stars and the Hercules Stream}
\label{sec:td}

The panels also show evidence of a prograde component at around
$J_\phi\approx 1100$, $J_R \approx 150$ km\,s$^{-1}$ kpc, $E \approx
-1.6$ km$^2$s$^{-2}$, which is present at all metallicities as a
overdense clump distinct from the general distribution.  The component
is marked by an ellipse in the panels of Fig.~\ref{fig:actspace}. We
fit a Gaussian with a flat background to isolate this substructure and
find a component with mean and dispersion $\langle J^{\rm
  C}_\phi\rangle \approx 1100$ km\,s$^{-1}$ kpc and $\sigma^{\rm
  C}_{J_\phi} \approx 320$ km\,s$^{-1}$ kpc.  It is comprised of stars
moving on disk-like prograde orbits with intermediate
energies. Surprisingly, these stars show a noticeable positive
$\langle v_R^{\rm C} \rangle \approx 12$ km\,s$^{-1}$, so that they have
a net outward motion in the Galactic rest frame. This is illustrated
in the leftmost panel of Fig.~\ref{fig:thirdcomp}. The positive
$\langle v_R \rangle$ signal comes mostly from low Galactic
latitudes. The low latitude stars between $|z| <3.5$ kpc (magenta) show
positive mean $\langle v_R \rangle$ signatures at the same $J_\phi$ range
that resembles the signal from the whole sample (black).
In contrast, the high latitude stars with $|z| > 3.5$ kpc (green)
show close to zero mean $\langle v_R \rangle$ across
the $J_\phi$ range. The magnitude of positive mean $\langle v_R \rangle$
is significant compared to the corresponding standard error.

We select stars with $\langle J^{\rm C}_\phi\rangle \pm 1.25
\sigma^{\rm C}_{J_\phi}$.  The remaining panels of
Fig.~\ref{fig:thirdcomp} show the $\langle v_R \rangle$ trend of this
selected group (red) and the rest (blue).  The solid and dashed lines
indicate the low and high latitude stars for each group (separation at
$|z| = 3.5$ kpc).  Notice the clear difference in the magnitude of
$\langle v_R \rangle$ between the selected stars and the rest in
various distributions. We also note that the signal drops by an order
of magnitude at or near the solar radius and thereafter is close to
zero. Interestingly, there is no clear metallicity dependency as the
signal remains at a similar magnitude across the metallicity range.
Taken together, these facts strongly suggest the source of the signal
is dynamical in origin, namely a resonance. This is the first time
  that a resonance has been identified in such metal-poor stars.  The
selected stars at low Galactic latitude have $(\langle v_R^{\rm
  S}\rangle ,\langle v_\phi^{\rm S}\rangle ,\langle v_z^{\rm
  S}\rangle) = (12.2,140.3,3.0)$ km\,s$^{-1}$.  These values are
indicative, as the selection is crude and probably blended with the
other halo populations. However, the value of the mean rotational
velocity $\langle v_\phi^{\rm S} \rangle$ suggests that this component
may be associated with the thick disk.

The outward mean radial velocity suggests an association with the
Hercules stream, which is also moving outward and is located interior
to the Solar circle. Stream is a slight misnomer, as the Hercules
stream is really a co-moving group of stars of dynamical origin,
induced probably by the Outer Lindblad Resonance of the Galactic
bar~\citep{De00}. The Hercules stream has a complex structure with
outward $\langle v_R \rangle$ somewhat larger than we
measure~\citep{An14,Hu18}, though this is likely accounted by
contamination in our sample. It has previously been detected in stars
with metallicities $-1.2<$ [Fe/H] $<0.4$~\citep{Be07}.  However, we
see from the panels in Fig.~\ref{fig:thirdcomp} that -- if our
  identification with the Hercules stream is correct -- then it is
detectable right down to [Fe/H] $\approx -2.9$ and so is present in
stars of metallicity normally associated with thin disk, thick disk
and halo.

\section{Discussion}

There are a number of possible explanations of the properties of the
Milky Way stellar halo in action space.  The highly flattened
metal-rich component is almost certainly the residue of the disruption
of accreted dwarf galaxies. The strong radial anisotropy already
suggests that the progenitors of this component fell in from large
distances. Infall of satellites with random alignments tends to
isotropize the dispersion tensor. The easiest way to maintain such
extreme radial anisotropy is through the infall of one satellite, or
group infall of multiple satellites, from a preferred direction
~\citep{Be18}. The eccentric halo substructure discovered by
\citet{My17} using the Tycho--{\it Gaia} Astrometric Solution (TGAS)
crossmatched with RAVE may well be part of this component.  The origin
of the rounder, metal-poor component is less clear. The isotropic
kinematics of this population, together with its roughly spherical
shape and mild prograde rotation, are reminiscent of the halo globular
clusters.  If these objects were once much more massive than the
entities surviving today~\citep{Sc11}, then they may have contributed
most of the stars in the metal-poor component of the
halo. Alternatively, the accretion of low mass dwarfs along random
directions may have built the metal-poor component. Another
contributor could be levitation~\citep{Sr96}. The growing thin disk
can trap stars in the 2:2 resonance and lift them to higher
latitude. A pre-existing structure of metal-poor stars could be
fattened by such a process.

In addition to these two well-known components, we have identified a
high energy subpopulation that is very strongly
retrograde. \citet{He17} noticed that a high fraction of stars more
loosely bound than the Sun are retrograde in the local TGAS
sample. Here, we have shown that these stars are overwhelmingly
metal-rich, and that the feature does not extend to stars with [Fe/H]
below $-1.9$.  A candidate for a retrograde invader exists in the
anomalous globular cluster $\omega$Centauri, long suspected to be the
nucleus of a stripped dwarf galaxy~\citep{Be03}. It is known to be on
a retrograde orbit.  \citet{Ma12} showed, on the basis of
chemodynamical evidence, that it is a major source of retrograde halo
stars in the inner Galaxy.  The progenitor of $\omega$Centauri has to
be massive, so that the satellite is dragged deep into the potential
of the Milky Way and placed on its present low energy orbit, which is
marked by a purple star on Fig.~\ref{fig:actspace}.  While
$\omega$Centauri is on an eccentric orbit, it is considerably
retrograde as well, given its present-day energy.  Estimates of its
initial mass are typically $\sim 10^{10} \Msun$ \citep{Ts03,Va11},
while its present-day mass is only $5 \times 10^6 \Msun$
\citep{Me95}. Its disruption must have sprayed high energy retrograde
stars throughout the inner Galaxy, even though it is on a low energy
orbit now.

Finally, action space has allowed us to trace a new, resonant
  component down to very low metallicities. This component is also
flattened and has prograde rotation with a mean of $\langle v_\phi
\rangle \approx 140$ km\,s$^{-1}$. Its most unusual feature is a
small, but statistically significant, outward mean radial velocity
($\langle v_R \rangle \approx 12$ km\,s$^{-1}$). This is present
across the swathe of metallicities in our sample, namely $-2.9<$
[Fe/H]$ < -1.3$, yet is spatially restricted to just within the Solar
circle and at $|z| <3.5$ kpc. The confinement of this prograde feature
to a range in Galactocentric radii is unusual and strongly supports a
resonance origin.  A link with the Hercules stream in disk
  stars~\citep[e.g.,][]{Hu18} seems likely.

There are a number of ways the forthcoming {\it Gaia} Data Release 2 (DR2)
can confirm our picture. First, the crossmatches with radial velocity
surveys will have well-defined selection functions. This will allow
use of halo distribution functions~\citep{Wi15,Po15} to separate the
populations in action space as a function of metallicity and hence
better characterise their properties.  Secondly, the SDSS--{\it Gaia}
catalogue undersamples low latitude stars and so coverage of such
substructures is patchy. With {\it Gaia} DR2, stars with kinematics akin to
the Hercules stream can be traced as a function of location throughout
the Galaxy. Thirdly, {\it Gaia} colours, as well as spectroscopic follow-up
to obtain alpha-abundances, may confirm the origin of the halo
components. For example, if the high energy, retrograde component
comes from $\omega$Centauri, the stripped stars in action space must
be trackable to their present position through chemodynamical
data. Fourthly, the improved proper motions may even permit the use of
angles (as opposed to actions) in studying resonances throughout the
halo.

\acknowledgments GCM thanks the Boustany Foundation, Cambridge
Commonwealth, European \& International Trust and Isaac Newton
Studentship for their support of his work.  JLS thanks the Science and
Technology Facilities Council for financial support. The research
leading to these results has received partial support from the
European Research Council under the European Union's Seventh Framework
Programme (FP/2007-2013) / ERC Grant Agreement no. 308024. This work
has made use of data from the European Space Agency (ESA) mission {\it
  Gaia} (\url{https://www.cosmos.esa.int/gaia}), processed by the {\it
  Gaia} Data Processing and Analysis Consortium (DPAC,
\url{https://www.cosmos.esa.int/web/gaia/dpac/consortium}). Funding
for the DPAC has been provided by national institutions, in
particular the institutions participating in the {\it Gaia}
Multilateral Agreement.

\end{document}